   \documentstyle[sprocl,epsf]{article}

\begin{document}

\title{
NUCLEAR EFFECTS IN THE $F_3$ STRUCTURE FUNCTION 
}

\author{S. A. KULAGIN}

\address{
        Institute for Nuclear Research,
        Russian Academy of Sciences,
        Moscow, Russia\\ E-mail: kulagin@ms2.inr.ac.ru
}

\maketitle\abstracts{
We discuss nuclear effects in the structure function $F_3$
measured in neutrino deep-inelastic scattering
on heavy nuclear targets.
}

%%%%%%%%%%%%%%%%%%%%%%%%%%%%%%%%%%%%%%%%%%%%%%%%%%%%%%%%%%%%%%%%%%%%%%%

Experiments on deep-inelastic scattering (DIS) of charged leptons and
neutrino remains an important source of information about the nucleon
as well as about the nuclear structure.
DIS experiments with neutrino beams \cite{CCFR:96,CSB97}
has reached now an accuracy comparable
to that of experiments with charged leptons.
Note in this respect that because of statistics reasons neutrino data
are collected for heavy nuclei like iron \cite{CCFR:96}.

Nuclear effects have been extensively discussed
for the
spin independent structure functions (SF) $F_{1,2}$ as well as for
the spin SF $g_{1,2}$ which are measured in
the charged lepton DIS
(for a recent reviews of experimental situation and theoretical
approaches to nuclear effects in DIS 
see refs.\cite{Arneodo:94,GST95,CoPr,BCK:92}),
however untill now
only a little attention was payed to
nuclear effects in neutrino DIS.\cite{SiTo95,Kumano,Kul98,BLT98}
It is usually assumed in analysing neutrino data
that nuclear corrections are the same
as in charged lepton DIS.
At large $Q^2$ this assumption can be motivated by the parton model
where the charged lepton and neutrino SF
are expressed in terms of universal parton distributions.
However the similarity between the charge lepton and neutrino $F_2$
fails at small $x$ where they
are different in the strange and charm
quarks content as well as when one studies effects due to finite $Q^2$.
In this respect the situation is even more uncertain with the SF $F_3$,
which does not have its analog in the charged lepton DIS.

In the present contribution we report on the results of our
studies of nuclear effects in the SF
$F_3$ within an approach applicable
at finite $Q^2$. More detailed discussion on this matter
can be found in ref.\cite{Kul98}.

Nuclear effects in DIS are usually discussed
within the context of the impulse approximation for the
nucleons, where the nuclear
Compton scattering amplitude is approximated by
the uncoherent sum of the scattering amplitudes from bound nucleons
neglecting final state interactions.
An argument to support this approximation comes from the analysis
of characteristic space-time scales involved in deep-inelastic
scattering.
In the laboratory frame a characteristic time for DIS is
$1/Mx$, where $M$ is the nucleon mass (see, e.g., ref.\cite{IoKhLi84}).
For $x>0.2$ this time is smaller than the typical average
distance between bound nucleons in the nucleus and the scattering
proceeds uncoherently.
Major nuclear effects in this region are due to
Fermi-motion and nuclear binding~\cite{fm},
and possible off-shell modification of nucleon structure functions
\cite{MeScTh94,KPW94,KMPW95} (for more discussion and references
see ref.\cite{GST95}).

On the other hand at small $x$ the coherent multiple scattering
effects should be important. It is known that for the SF $F_2$ these
effects cause the nuclear shadowing phenomenon (for a review of
models for the nuclear shadowing see refs.\cite{CoPr,Arneodo:94,BCK:92}).
We note however that not much is known yet about the shadowing effect
in the SF $F_3$ \cite{Kumano}
%%\footnote{%
(the preliminary results of ref.\cite{KuMe}
show that the nuclear shadowing effect in
$F_3$ is even more prominent than the one in $F_2$).

We comment also that correction due to meson exchange currents in $F_3$
vanishes, because light mesons, such as $\pi$ and $\rho$,
have vanishing $F_3$ structure functions
(if one neglects a small contributions due to $s-\bar c$ and $\bar s-c$).

Based on these observations we consider nuclear Compton amplitude in
the impulse approximation. In order to find a connection
between the nuclear and nucleon SF one has to take the imaginary part of the
Compton amplitude and then extract the desired SF.
We note that there are few subtle points in this procedure,
which are usually neglected in many calculations based on
convolution model.
First of all we observe that bound nucleons are off-mass-shell
($p^2\not= M^2$) and nuclear SF
are not determined by those of on-mass-shell
nucleon but are sensitive to their off-shell behavior.
As it is discussed in \cite{MeScTh94,KPW94,KMPW95,Kul98},
the Lorentz-Dirac structure of
the nucleon hadronic tensor in off-shell region is more complicated
than that of the on-shell one.
A direct reason for this is that one can not use the
Dirac equation for the off-shell nucleon which greatly reduces the
number of independent amplitudes in the hadronic tensor.
In particular we found that the SF $F_3$ splits into
four independent inelastic form-factors for the nucleon off-shell.
Similar observations \cite{KPW94,KMPW95}
have been already done for the SF
$F_2$ as well as for $g_1$ and $g_2$.
One should conclude therefore that even in the impulse approximation
there is no simple factorization between
the nuclear and nucleon SF, that is assumed in convolution model
calculations.

%Obviously these observations make the problem of extraction of
%the proton and neuteron SF from nuclear data
%more complicated and even could
%introduce a principal uncertainty into this problem.

%% 1/M EXPANSION AND NONRELATIVISTIC LIMIT
One can show however that
things simplify considerably in the limit
of weak nuclear binding \cite{Ku89,KPW94,KMPW95,Kul98}.
We have done
a systematic $1/M$-expansion of nuclear matrix elements which enter
the nuclear Compton amplitude and
found that the factorization for the SF is
remarkably recovered if one keeps only terms to order $1/M^2$
(including the latter).
In this approximation
the nuclear $F_3$ can be expressed in
terms of the generalized convolution of the (non-relativistic)
nuclear spectral function ${\cal P}(\varepsilon,{\bf p})$
and a combination of the four
off-shell nucleon inelastic form factors which we take for the
definition of the ``off-shell nucleon structure function"
$F_3(x,Q^2;p^2)$ (notice the dependce on the nucleon ``virtuality''
$p^2$ as an additional variable). The final result reads:
\begin{eqnarray}
\label{IAnr}
x\,F_3^A(x,Q^2) =
\sum_{\tau=p,n}\int{d\varepsilon d^3p\over (2\pi)^4}
{\cal P}^\tau(\varepsilon,{\bf p})
        \left(1+\frac{p_z}{{\gamma} M}\right)
x'\,F_3^\tau(x',Q^2;p^2) ,
\end{eqnarray}
where the integration is done over the nucleon
four-momentum,
$p=(M+\varepsilon,{\bf p})$,
 $x'=Q^2/2p\cdot q$
is the Bjorken variable of the bound nucleon
and $\gamma=|{\bf q}|/q_0$ is the ratio of the space to time components
of the momentum transfer.
We note that no
approximation is done with respect to $Q^2$, so that Eq.(\ref{IAnr})
is valid, in general, for any $Q^2$.

We use Eq.(\ref{IAnr}) to separate $Q^2$ dependence of the nuclear
SF due to nuclear effects.
A useful observation is that these effects come through
dependence on $\gamma$ of the variable $x'$ as well as a ``flux"
factor in Eq.(\ref{IAnr}).
Making use of this observation we expand Eq.(\ref{IAnr})
in nuclear $Q^{-2}$ series:
\begin{eqnarray}
\label{q2-exp}
xF_3^A(x,Q^2)/A &\approx&
\left\langle\left(1+{p_z\over M}\right)
        x'F_3(x',Q^2;p^2)\right\rangle \nonumber\\
&&- {2M^2x^2\over Q^2}
        \left\langle {p_z\over M}{\partial\over\partial x'}
        \left(x'^2F_3(x',Q^2;p^2)\right)
        \right\rangle  + \cdots
\end{eqnarray}
Here to simplify the notations the brackets denote
the averaging over the nuclear spectral function.
Note that $x'$ in Eq.(\ref{q2-exp})
corresponds to the light-cone kinematics when
$\gamma=1$.
The dots denote terms of order
$Q^{-4}$ and higher which are not written here explicitly (for more
details see ref.\cite{Kul98}).

Few comments on Eq.(\ref{q2-exp}) are in order.  First we note that
Eq.(\ref{q2-exp}) makes it possible to separate the effects due to
$Q^2$ dependence of the nucleon SF itself and those which are due to
nuclear effects.  It was argued in ref.\cite{Kul98} that an effective
parameter in this expansion is $p_{\rm char}^2x^2/Q^2$ with $p_{\rm
char}$ being the characteristic momentum of the bound nucleon.  This
allows us to apply the $Q^{-2}$ expansion in Eq.(\ref{q2-exp})
down to small $Q^2\sim
0.1\,$GeV$^2$ and keep only $Q^{-2}$ correction at
practically interesting $Q^2$.
Fig.1 shows the $x$- and $Q^2$-dependence of the ratio
$R_3=\frac1A F_3^A(x,Q^2)/F_3^N(x,Q^2)$
calculated for the ${}^{56}$Fe nucleus assuming the latter to be an isoscalar
target with $N=Z$.

Eq.(\ref{q2-exp}) is particularly useful in calculating nuclear
corrections to the Gross-Llewellyn-Smith sum rule~\cite{GLS69}.
Nuclear corrections to the GLS sum rule cancel out
in the leading order, which is due to the baryon charge
conservation in strong interaction.
The $Q^{-2}$ correction to the GLS sum rule
turned out to be negative and small.
For the iron and deuterium nuclei we have the following estimates
for the corrections:
$\delta S_{\rm GLS}^{\rm Fe} = - 1.2\cdot10^{-2}/ Q^2,\ 
\delta S_{\rm GLS}^{\rm D} = - 1.9\cdot10^{-3}/ Q^2$, where the
coefficients are taken in the units of GeV$^2$.

Notice however that nuclear corrections are sizable for 
the GLS integrals truncated from below,
$S_{\rm GLS}(x,Q^2)=\int_x^1 dx' F_3(x',Q^2)$,
as it is illustrated in Fig. 2,
that may have an impact on extraction of the value
of the nucleon GLS sum rule from the data.

This work is supported in part by the RFBR grant 96-02-18897.
The author is grateful to the organizers of the DIS'98 Workshop
for warm hospitality.

\begin{figure}[htb]
 \epsfbox{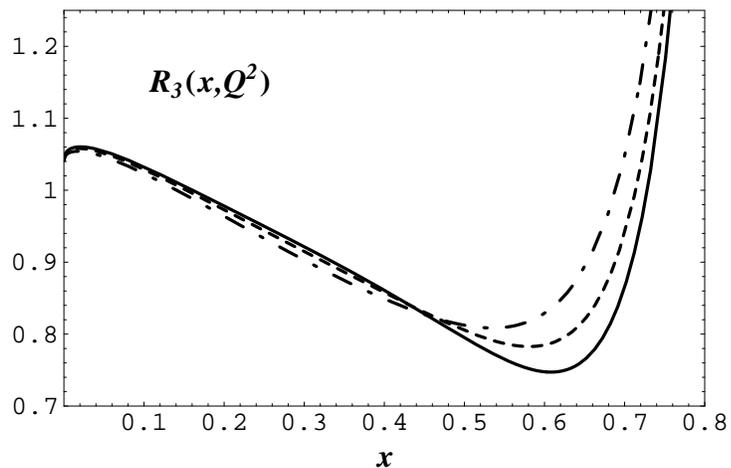}   %to be used with epsf.sty
\caption{Shown is the $Q^2$ dependence of the ratio
$R_3$ for the iron nucleus.
The dashed-dotted line corresponds to $Q^2=15\,$GeV$^2$,
the dashed line --- $Q^2=5\,$GeV$^2$,
while the solid line --- $Q^2=3\,$GeV$^2$.
}
\label{fig_ratio_q}
\end{figure}
\begin{figure}[htb]
 \epsfbox{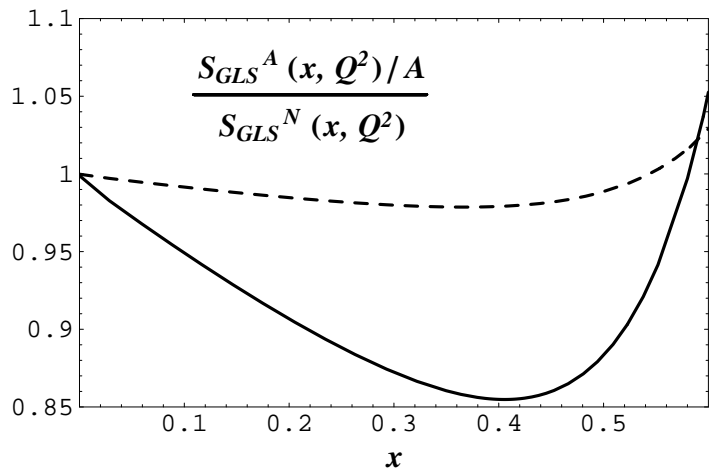}   %to be used with epsf.sty
\caption{Shown is the $x$ dependence of
the nucleus/nucleon ratio of the GLS integrals
calculated at $Q^2=5\,$GeV$^2$.
The dashed line corresponds to the deuteron nucleus,
while the solid line to the iron nucleus.
}
\label{fig_GLS_x}
\end{figure}

\end{document}